\documentclass[11pt]{article}
\usepackage{graphicx,amsmath}
\usepackage{cite}

\input{pubboard/babarsym}
\def\Dp      {\ensuremath{D^+}}
\def\Dm      {\ensuremath{D^-}}
\def\Dstar   {\ensuremath{D^{*}}}
\def\Dstarp  {\ensuremath{D^{*+}}}
\def\Dstarm  {\ensuremath{D^{*-}}}

\def\Abar    {\kern 0.18em\overline{\kern -0.18em A}{}\xspace}

\newcommand{\etal}{{\it et al.}}

\def\fish    {\ensuremath{\cal F}}

\def\acp {\ensuremath{{\cal A}}}

\def\thsph    {\ensuremath{\theta_{\scriptscriptstyle S}}}

\def\pipi     {\ensuremath{\pi\pi}}

\def\fish    {\ensuremath{\cal F}}
\def\cerenkov{Cherenkov}

\newcommand{\BABARPubYear}    {02}

\newcommand{\BABARConfNumber} {13}
\newcommand{\SLACPubNumber} {9304}

\long\def\inst#1{\par\nobreak\kern 4pt\nobreak
    {\it #1}\par\vskip 10pt plus 3pt minus 3pt}

\begin{document}
\pagestyle{empty}

\begin{flushright}
\babar-CONF-\BABARPubYear/\BABARConfNumber \\
SLAC-PUB-\SLACPubNumber \\
%hep-ex/\LANLNumber \\
July 2002 \\
\end{flushright}

\par\vskip 3cm

\begin{center}
{
\Large \bf
\boldmath
Measurements of Branching Fractions and Direct $\CP$ Asymmetries in $\pip\piz$, $\Kp\piz$
and $\Kz\piz$ \boldmath{$\B$} Decays
\unboldmath
}
\end{center}
\bigskip

\begin{center}
\large The \babar\ Collaboration\\
\mbox{ }\\
%\today
July 24, 2002
\end{center}
\bigskip \smallskip

% Abstract
\begin{center}
\large \bf Abstract
\end{center}
We present preliminary results of the analyses of $\Bp \to h^+ \piz$ 
(with $h^+$ = $\pip, \Kp$) and $\Bz \to \Kz \piz$ decays
from a sample of approximately $88$ million \BB\ pairs collected by the \babar\ detector 
at the \pep2\ asymmetric-energy $B$ Factory at SLAC.
We measure the $\pip\piz$ branching fraction and we obtain
\[
{\BR} (\Bp \to \pip \pi^0) = (5.5 ^{+1.0}_{-0.9} \pm 0.6) \times 10^{-6}
\]
with a significance of $7.7 \sigma$ including systematic uncertainties.
We measure the $\Kp\piz$ and $\Kz\piz$ branching fractions to be
${\BR}(\Bp \to \Kp \piz) = (12.8 ^{+1.2}_{-1.1} \pm 1.0) \times 10^{-6}$
and
${\BR}(B^0 \rightarrow \Kz \piz) = (10.4 \pm 1.5 \pm 0.8) \times 10^{-6}$.
At the same time, the direct \CP-violating asymmetries are investigated
and we find $\acp_{\pip\pi^0} = -0.03 _{-0.17}^{+0.18} \pm 0.02$,
$\acp_{\Kp\pi^0} = -0.09 \pm 0.09 \pm 0.01$ and $\acp_{\Kz \piz} =
0.03 \pm 0.36 \pm 0.09$, where the errors are statistical and systematic,
respectively.

\vspace{1.0cm}
\vfill
\begin{center}
Contributed to the $31^{st}$ International Conference of High Energy Physics,
\\7/24-7/31/2002, Amsterdam, The Netherlands
\end{center}

\vspace{1.0cm}
\begin{center}
{\em Stanford Linear Accelerator Center, Stanford University,
Stanford, CA 94309} \\ \vspace{0.1cm}\hrule\vspace{0.1cm} Work
supported in part by Department of Energy contract DE-AC03-76SF00515.
\end{center}

\newpage
\pagestyle{plain}

% Input author list file
\begin{center}
\small

The \babar\ Collaboration,
\bigskip

%% author list as of 05-Jul-2002 (556 authors)
B.~Aubert,
D.~Boutigny,
J.-M.~Gaillard,
A.~Hicheur,
Y.~Karyotakis,
J.~P.~Lees,
P.~Robbe,
V.~Tisserand,
A.~Zghiche
\inst{Laboratoire de Physique des Particules, F-74941 Annecy-le-Vieux, France }
A.~Palano,
A.~Pompili
\inst{Universit\`a di Bari, Dipartimento di Fisica and INFN, I-70126 Bari, Italy }
% ...

\end{center}\newpage
 \setcounter{footnote}{0}
\section {Introduction}
The study of $B$ meson decays into charmless hadronic final states
plays an important role in the understanding of \CP\ violation in the
$B$ system.  Measurements of the \CP-violating asymmetry in the $\pip\pim$
decay mode can provide information on the angle $\alpha$ of the
Unitarity Triangle. However, in contrast to the theoretically clean
determination of the angle $\beta$ in $B$ decays to charmonium final
states \cite{sin2betaBaBar,sin2betaBelle} the  extraction
of $\alpha$ from $\pip\pim$ decays is complicated by the interference of
$b\to uW^-$ tree and $b\to dg$ penguin amplitudes. Since these
amplitudes have similar magnitude but carry different weak phases, 
additional measurements of the isospin-related decays\footnote{Charge
conjugate modes are assumed throughout this paper unless explicitly stated.},
$\Bp \to \pip \piz$ and $\Bz \to \piz \piz$, are required to provide a
way of measuring the angle $\alpha$~\cite{ref:gronau2}. The measurement
of the branching fraction of the $\Bp \to \pip \piz$ decay is, in fact, a
crucial ingredient, since it is a pure tree decay to a very
good approximation. Therefore, in this channel direct \CP\ violation,
detected as a charge asymmetry ($\acp_{\pip\piz}$), is expected to be zero. 

On the other hand, measurements of $\B \to K\pi$ decays
can be related to a model dependent extraction of 
the weak phase $\gamma$ with a global fit to the observables.
In order to do this, several models have been 
proposed, based on different dynamical assumptions for 
$B$ decays to two light pseudoscalar mesons
\cite{beneke,ciuchini,Keum:2000wi,Isola:2001bn}. 
All these approaches are able to reproduce experimental values for
branching fractions,
but they do not show a good sensitivity to the value of the weak phase 
$\gamma$.
Providing information on \CP-violating asymmetries could contribute to
increase this sensitivity, clarifying the theoretical framework and
improving our ability to constrain the Unitarity Triangle in the
$(\bar \rho, \bar \eta)$ plane~\cite{ckm-w}.

In the case of $\Kz\piz$, the ideal measurement is a time dependent
analysis of the $\KS\piz$ final state, which is a $\CP$ eigenstate. In this
way, one would be sensitive to the direct \CP-violating asymmetry
(related to the coefficient of the cosine term) and to the weak
phase entering the decay (related to the coefficient of the sine
term)~\cite{tag}. In the present analysis, we perform a time
integrated study, providing the direct \CP-violating asymmetry
defined as
\[
\acp_{\CP} = \frac{\vert \Abar \vert ^2-\vert A \vert ^2}
{\vert \Abar \vert ^2+\vert A \vert ^2},
\]
where $A$ ($\Abar$) is the decay amplitude (its $\CP$ conjugated)
taken into account.
We actually investigate $\acp_{\KS\piz}$ which, in the SM, is equal to
$\acp_{\Kz\piz}$, neglecting contributions with more than one weak
boson.

We present here results on the $\Bp \to \pip\piz$,
$\Bp \to \Kp\piz$ and $\B^0 \to \Kz \piz$ decays.
The \babar\ collaboration has previously published \cite{ourPRL}
observations of these channels: we now have reduced
the errors on branching fractions and investigated direct \CP-violating
effects.

\section {The \babar\ Detector and Data Set}
The data used in these analyses were collected with the \babar\ detector
at the \pep2\ $\epem$ storage ring.  
The sample corresponds to an integrated luminosity of  
about $81\invfb$ accumulated on the
\FourS\ resonance (``on-resonance'') and about $9 \invfb$ accumulated at a 
center-of-mass (CM) energy about $40\mev$ below the \FourS\ resonance
(``off-resonance''), which are used for background studies.
The on-resonance sample corresponds to $(87.9 \pm 1.0)\times 10^6$ 
\BB\ pairs.  The collider is operated with asymmetric beam energies, 
producing a boost ($\beta\gamma = 0.55$) of the \FourS\ along the 
collision axis.  

\babar\ is a solenoidal detector optimized for the asymmetric beam
configuration at PEP-II and is described in detail in Ref.~\cite{babarnim}.
Charged  particle (track) momenta are measured in a tracking system
consisting of a 5-layer, double-sided, silicon vertex tracker and a
40-layer drift chamber filled with a gas mixture of helium and
isobutane, both operating within a $1.5\,{\rm T}$ superconducting
solenoidal magnet. Photon candidates are selected as local maxima of
deposited energy in an electromagnetic calorimeter (EMC) consisting
of 6580 CsI(Tl) crystals arranged in barrel and forward endcap
subdetectors. In this analysis, tracks are identified as pions or
kaons by the \cerenkov\ angle $\theta_c$ measured by a detector of
internally reflected \cerenkov\ light (DIRC). The DIRC system is a
unique type of \cerenkov\ detector that relies on total internal
reflection within the radiating volumes (quartz bars) to deliver the 
\cerenkov\ light outside the tracking and magnetic volumes,
where the \cerenkov\ ring is imaged by an array of $\sim 11000$
photomultiplier tubes.
The \cerenkov\ angle $\theta_c$ is measured with a typical resolution
of $3$ mrad, with a separation between kaons and pions of
$8\sigma$ ($2.5\sigma$) at $2\gevc$ ($4\gevc$).
Good separation at high momenta is essential for two-body $B$ decay
since the boost increases the momentum range of the decay products
from a narrow distribution centered near $2.6\gevc$ in the CM to a
broad distribution extending from $1.7$ up to $4.3\gevc$.

\section {Event Selection,  \boldmath{$\piz$}  and  \boldmath{$\Kz$} 
Reconstruction}
Hadronic events are selected based on track multiplicity and event
topology. Backgrounds from non-hadronic events are reduced by
requiring the ratio of the second to zeroth Fox-Wolfram moment~\cite{fox}
to be less than $0.95$ and the sphericity~\cite{spheric} of the event
to be greater than $0.01$.

Charged $\pi$ and $K$ candidates (except for $\KS$ daughters) are
reconstructed within the tracking fiducial volume and quality criteria are imposed:
they are required to originate within $1.5$~cm in the $xy$ plane
and $10$~cm in $z$ from the interaction point, to have at least $12$ measured
drift chamber hits and a minimum transverse momentum of 0.1~GeV/$c$.

Candidate \piz\ mesons are reconstructed as pairs of photons
with an invariant mass within $3 \sigma$ of the nominal \piz\ mass~\cite{PDG},
where the resolution $\sigma$ is about 8 \mevcc.  
Photon candidates are selected as showers in the EMC that have
the expected lateral shape, are not matched to a charged track, and have a
minimum energy of 30 \mev. The \piz\ candidates are then kinematically
fitted with their mass constrained to the \piz\ nominal mass.

\Kz\ mesons are detected in the mode $\Kz\to \KS\to \pip\pim$. $\KS$
candidates are reconstructed from pairs of oppositely charged tracks that
form a well-measured vertex and have an invariant mass within $11.2\mevcc$
(which corresponds to $3.5\sigma$) of the nominal \KS\ mass~\cite{PDG}. 
The measured proper decay time of the \KS\ candidate is required
to exceed $5$ times its uncertainty.

\section {\boldmath{$B$} Reconstruction }
\label{brecosec}
Charged (neutral) $B$ meson candidates are reconstructed by combining a 
\piz  candidate with a track $h^+$ (a $\KS$ candidate). 
The kinematic constraints provided by the
\FourS\ initial state and  knowledge of the beam
energies are exploited to efficiently identify $B$ candidates.   We
define a beam-energy substituted mass  $\mes = \sqrt{E^2_{\rm
b}-\mathbf{p}_B^2}$, where $E_{\rm b} =(s/2 + \mathbf{p}_i
\cdot\mathbf{p}_B)/E_i$, $\sqrt{s}$ and $E_i$ are the total energies
of the \epem\ system in the CM and lab frames, respectively, and
$\mathbf{p}_i$ and $\mathbf{p}_B$ are  the momentum vectors in the lab
frame of the \epem\ system and the $B$ candidate, respectively. 
An additional kinematic parameter $\Delta E$  is defined as the
difference between the energy of the $B$ candidate and half the energy
of the \epem\ system, computed in the CM system. 
The \mes\ resolution is dominated by the beam energy spread, while  
for $\Delta E$ the main contribution comes from the measurement of 
particle energies in the detector. These two variables are therefore 
substantially uncorrelated ($\leq 6\%$).

For all the decay modes, the signal energy-substituted mass is parameterized
on simulated signal events by a Crystal Ball\footnote{The Crystal Ball
function~\cite{cryball} is a core Gaussian with a power law to describe
the left tail.} function.
The resolution is found to be about 2.9 \mevcc and it is
validated by comparing data and Monte Carlo resolutions for decays
into open charm final states with large branching fractions,
such as $\Bm \to D^0 \rho^{-}$, (with $\rho^{-} \to \pim \piz$ and
$D^0 \rightarrow K^- \pi^+$).

The $\Delta E$ distribution for the signal $h^+ \piz$
and $\Kz \piz$ events is described by another Crystal Ball function.
The mean value of this distribution is directly obtained from the fit
for the $h^+\piz$ sample and assumed to be the same for $\Kz \piz$
since the dominant effect comes from the $\piz$ energy scale.
For $\Kp \piz$ candidates, the mean of $\Delta E$ is also shifted because
the pion mass is assumed for all charged tracks in order to extract
the yields of both modes and the $\Delta E$ mean from one fit.
This shifted mean value can be expressed as
\begin{equation}
\nonumber
\langle \Delta E \rangle = -\gamma_{\rm boost}\times
\left( \sqrt{M_K^2+p^2}-\sqrt{M_\pi^2+p^2} \right)\, ,
\end{equation}
where $p$ is the momentum of the assumed kaon track, and $M_\pi$ and $M_K$
are the pion and kaon mass values, respectively. We estimate the resolution
on $\Delta E$ to be about $42$ \mev, based on simulated $\Bp \to h^+ \piz$ and
$ \B^0 \to \KS \piz$ events and cross-checked on the $\Bm \to D^0 \rho^{-}$
control sample.

Candidates are selected in the range $5.2<\mes<5.3\gevcc$. 
Different requirements on $\Delta E$ specific to each analysis
are described later.

\section { Background Rejection}
The dominant background to these channels is the continuum background,
coming from random combinations of a true \piz with a track (a true \KS), 
produced in $\epem\to \qqbar$ continuum events (where $q=u$, $d$, $s$ or $c$). 
Another source of background originates from $B$ decays into other
light charmless meson final states (charmless background).
The main contribution to this background comes from
$\Bp \to \rho^+ \piz$ and $B^{0} \to \rho^{\pm} \pi^{\mp}$
in the case of $\Bp \to \pip \piz$ and 
$\Bp \to K^{* +} \piz$ in the case of 
$\Bp \to \Kp \piz$ and $B^{0} \to \Kz \piz$.
Detailed Monte Carlo simulation, off-resonance, and on-resonance data  
are used to study backgrounds.

In the CM frame the continuum background typically exhibits a
two-jet structure, in contrast to the isotropic decay of $\BB$ pairs
produced in \FourS decays. We exploit the topology difference between
signal and background by making use of two event-shape quantities.
The first variable is the angle $\thsph$ between the sphericity axes
of the $B$ candidate and of the remaining tracks and photons in the
event. The distribution of $|\cos\thsph|$ in the CM frame is strongly
peaked near $1$ for continuum events and is approximately uniform for
\BB\ events. We require $|\cos\thsph| < 0.8$.
The second quantity is a Fisher discriminant
${\cal F}$, constructed from the quantities
$L_0$ and $L_2$, where $L_j$ is:
\begin{equation}
\nonumber
L_j=\sum_i p_i |\cos{\theta_i}|^j
\label{Eq:legendre}
\end{equation}
and $p_i$ ($\theta_i$) is the momentum (the angle with respect to the
thrust axis of the $B$ candidate in the CM frame) of each charged track
and neutral cluster not used to reconstruct the candidate $B$ meson.
Monte Carlo samples are used to obtain the values of the Fisher
coefficients, which are determined by maximizing the statistical
separation between signal and background events. 
No requirement is applied on ${\cal F}$; instead the distributions for
signal and background events are included in a maximum likelihood fit as
described in the next section.

On the other hand, charmless background events tend to peak in \mes,
as do signal events, and have the same ${\cal F}$ distribution as signal, 
since they are true $B$ decays. Nevertheless they are characterized by 
lower $\Delta E$ values, since at least a pion is lost in the
$B$ reconstruction. A cut on $\Delta E$ is the only way to
reduce this background to a negligible level.
We use on-resonance data in the negative $\Delta E$ sideband region
($-0.40 < \Delta E < -0.20\gev$)
to estimate the magnitude of this background.
The efficiency determined from Monte Carlo events is used to scale
the number of events in the sidebands and get the expected number
of events in the chosen signal region.
We finally require $-0.11 < \Delta E  < 0.15 \gev$ for $h^+\piz$ 
and $-0.15 < \Delta E < 0.15 \gev$ for $\Kz \piz$, reducing
the charmless background contribution to the level of less than $1\%$.

A total of $21752$ candidates in the on-resonance data satisfy the
$h^+\piz$ selection, and a total of $2668$ candidates satisfy the 
$\KS\piz$ selection. These two samples enter into two separate maximum
likelihood fits.

The final signal selection efficiency $\epsilon$ is $(26.1 \pm 1.7)\%$
for $\Bp \to \pip \piz$ and $(21.5 \pm 1.5)\%$ for $\Bp \to \Kp \piz$
events, while it is $(28.0 \pm 2.0)\%$ for $\Bz \to \KS \piz$ events.
The errors on the efficiencies are statistical and systematic, 
combined in quadrature. The dominant contribution to the systematic error
is due to the imperfect knowledge of \piz\ and \KS\ reconstruction
efficiencies ($5\%$ and $3\%$ relative errors, respectively).
The hierarchy of efficiency values comes from the difference in $\Delta E$
lower cut ($h^+\piz$ vs. $\KS\piz$) and from the use of the $\pi$ mass
hypothesis in the $\Delta E$ calculation ($\pip\piz$ vs. $\Kp\piz$).

\section{Signal Extraction}
For each topology ($h^+\piz$ and $\Kz\piz$), an unbinned maximum likelihood fit
determines the signal and background yields $n_i$
($i=1$ to $M$, where $M$ is the total number of signal and background types)
and $\CP$ asymmetries.
The measured asymmetry is defined as:
\[
\acp_m^i =\frac{ \bar n_i - n_i }{ \bar n_i + n_i }
\]
where $\bar n_i$ is the fitted number of $i^{th}$ type $h^{-} \piz$
[$\Kzb \piz$] events and $n_i$ corresponds to
$h^{+} \piz$ [$\Kz\piz$] events.
The input variables ($\vec{x}_j$) to the fit are \mes, $\Delta E$,
\fish\ and, in the case of charged modes, the \cerenkov\ angle
$\theta_c$ of the track from the candidate $B$ decay to distinguish
between the final states with $h=\pi$ and $h=K$.

The $h^+\piz$ extended likelihood function $\cal L$ is defined as
\begin{equation}
\nonumber
{\cal L}= \exp\left(-\sum_{i=1}^M n_i\right)\, \prod_{j=1}^N
\left[\sum_{i=1}^M \frac{1}{2}(1- q_j \acp_m^i)   n_i 
{\cal P}_i\left ( \vec{x}_j; \vec{\alpha}_i\right) \right]\, ,
\end{equation}
where $q_j$ is the charge of the track $h$ in the $j^{th}$ event and,
in this case, $M$ is equal to $4$ including signal and background
$\pip\piz$ and $K^+\piz$.
The $M$ probabilities ${\cal P}_i(\vec{x}_j;\vec{\alpha}_i)$ are
evaluated as the product of probability density functions (PDFs) for
each of the independent variables $\vec{x}_j$, given the set of
parameters $\vec{\alpha}_i$ which define the PDF shapes.
Monte Carlo simulation is used to validate the assumption that the fit
variables are uncorrelated. The exponential factor in the likelihood
accounts for Poisson fluctuations in the total number of observed
events $N$.

For the $\Kz\piz$ mode we need to measure the flavor of the $B$ candidate
in order to extract the $\CP$ asymmetry. We use $B$ flavor tagging information
to distinguish $\Bz$ from $\Bzb$ decays.
$B$ candidates are defined $\Bz$ when the other $B$ is
recognized to be a $\Bzb$ and vice versa. Details of $\babar$ $B$ flavor tagging can
be found in Ref.~\cite{tag}. We have four different tagging categories ($k=1,..,4$), 
with different tagging efficiencies $\epsilon^k$ and wrong tag fractions $w^k$.
We also use untagged events ($k=0$) in the fit.
The extended likelihood is 
\begin{equation}
\nonumber
{\cal L}= \exp\left(-\sum_{i=1}^M n_i\right)\, \prod_{j=1}^N
\left[\sum _{k=0}^4\sum_{i=1}^M \frac{1}{2}\delta(c_j-k)(1- f_j  (1-2w^k)
\acp_m^i)  
n^k_{i} {\cal P}_{i}^k\left(\vec{x}_j; \vec{\alpha}_{i_k}\right) \right]\, ,
\end{equation}
where $c_j$ and $f_j$ are the tagging category and the measured flavor of
the $j^{th}$ event, $n_{i}^k=\epsilon^k n_{i}$ is the number of events of
the $i^{th}$ type in the $k^{th}$ category and $M$, in this case,
is equal to $2$ including signal and background.
The ${\cal P}_{i}^k(\vec{x}_j; \vec{\alpha}_{i_k})$ are
in principle category dependent. We found that such dependence can be
ignored without any bias in the fit, for all the PDFs, except for background
\mes\ and background $\fish$.

The \mes\ shape in background is parameterized by
the ARGUS threshold function~\cite{argus}
\begin{equation}
f(x)\propto x\sqrt{1-x^2}\exp[-\xi(1-x^2)]\,
\nonumber 
\end{equation}
where $x=\mes/m_0$ and $m_0$ is the average CM beam energy, and
$\xi$ is the parameter determining the shape of the distribution
and is left free to float in the likelihood fits. The background shape in
$\Delta E$ is parameterized as a second-order polynomial whose coefficients
are taken from a fit to the on-resonance \mes\ sideband region.
The signal distributions have been already described in Sect.~\ref{brecosec}.

Events from simulated signal decays and from on-resonance 
\mes\ sideband regions are used to parameterize the Fisher discriminant 
PDFs for signal and background events as a bifurcated Gaussian\footnote{The
bifurcated Gaussian is an asymmetric Gaussian having two different $\sigma$,
one for $x > \mu$ and another for $x < \mu$, where $\mu$ is the mean value.}
and a sum of two Gaussians, respectively. Alternative parameterizations for 
${\cal F}$,
obtained from off-resonance data (for background) and from a large sample of
fully reconstructed $\Bz \to D^{(*)}n\pi$ decays (for signal), are used to 
estimate systematic uncertainties. The $\theta_c$ PDFs are derived from kaon
and pion tracks in the momentum range of interest from a sample 
of $D^{*+}\to\Dz\pip$ ($\Dz\to \Km\pip$) decays. This
control sample is used to parameterize the $\theta_c$ resolution ($\sigma_{\theta_c}$)
as a function of track polar angle; double-Gaussian PDFs are constructed
from the difference between measured and expected values of $\theta_c$
for the pion or kaon hypothesis, normalized by the error $\sigma_{\theta_c}$.
Tagging efficiencies and mistag fractions are estimated from a sample
of fully reconstructed $B$ decays.

The results of the fit are summarized in the first column of Table
\ref{tab:hpi0brresult}, where the statistical error for each mode
corresponds to a $68\%$ confidence level interval and is given by the
change in signal yield $n_i$ that corresponds to a $-2\ln{\cal L}$
increase of one unit.
In the case of the $\pip \piz$ final state, we evaluate how the
imperfect knowledge of the PDF shapes can affect the significance of
the signal. We recalculate the square root of the change in
$-2\ln{\cal L}$ with $n_{\pip\piz}$ fixed to zero for the worst case
PDF variations and we find a statistical
significance of $7.7 \sigma$  for the signal.

\begin{figure}[!tbp]
\begin{center}
\begin{minipage}[h]{6.0cm}
  \includegraphics[width=6.0cm]{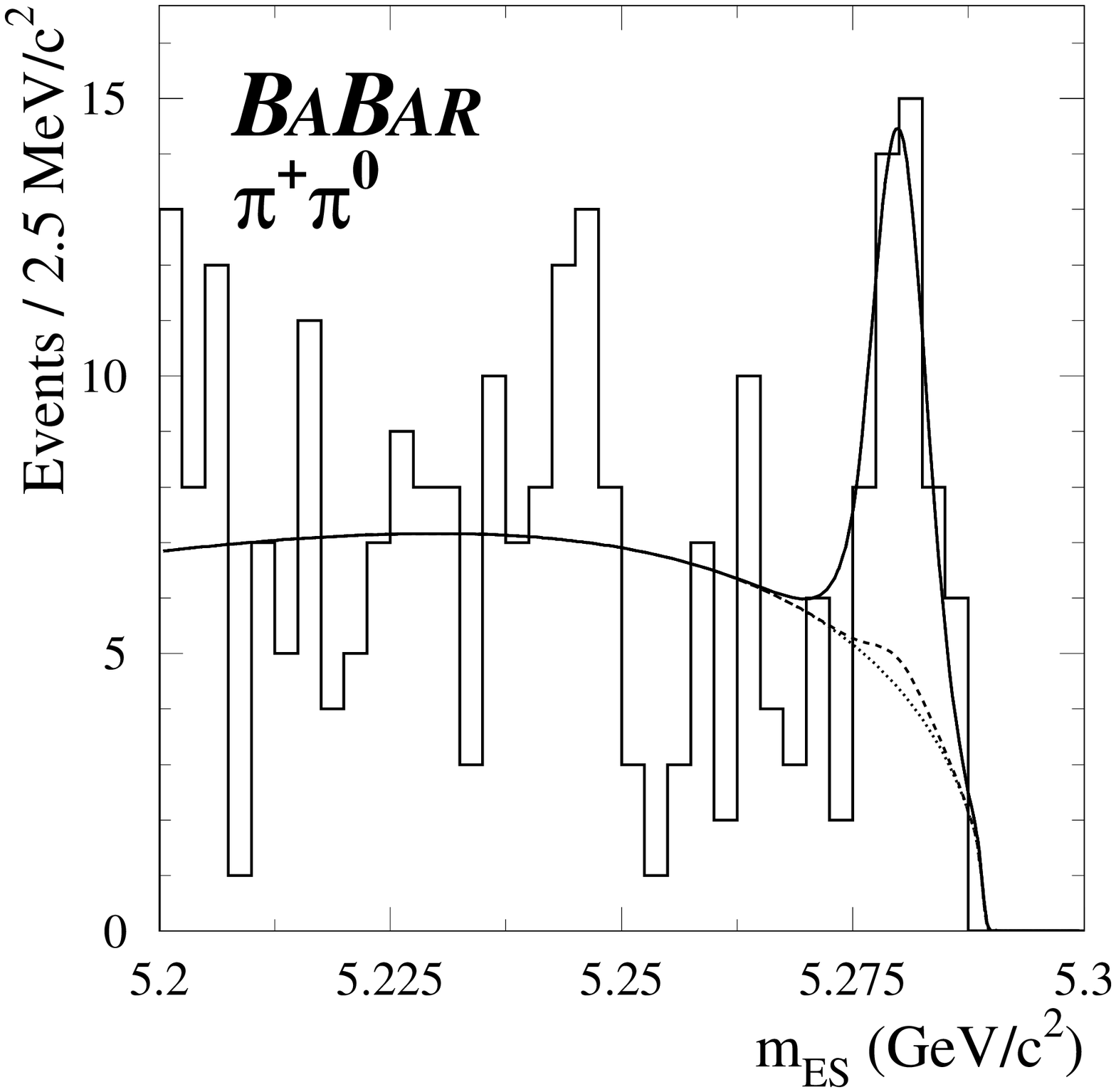}
\end{minipage}
\begin{minipage}[h]{6.0cm}
 \includegraphics[width=6.0cm]{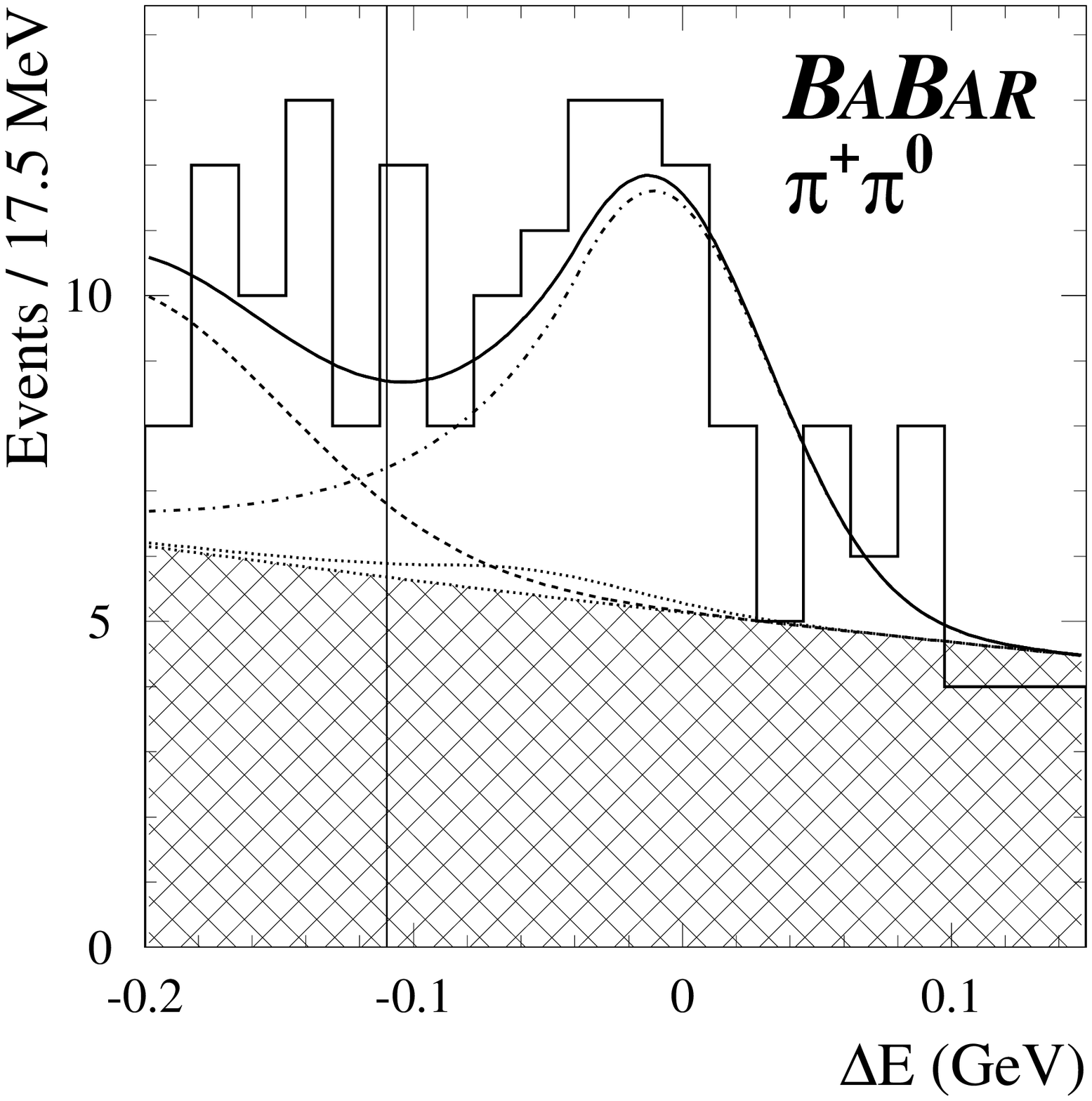}
\end{minipage}
\end{center}
\vspace{-0.9cm}
\begin{center}
\begin{minipage}[h]{6.0cm}
  \includegraphics[width=6.0cm]{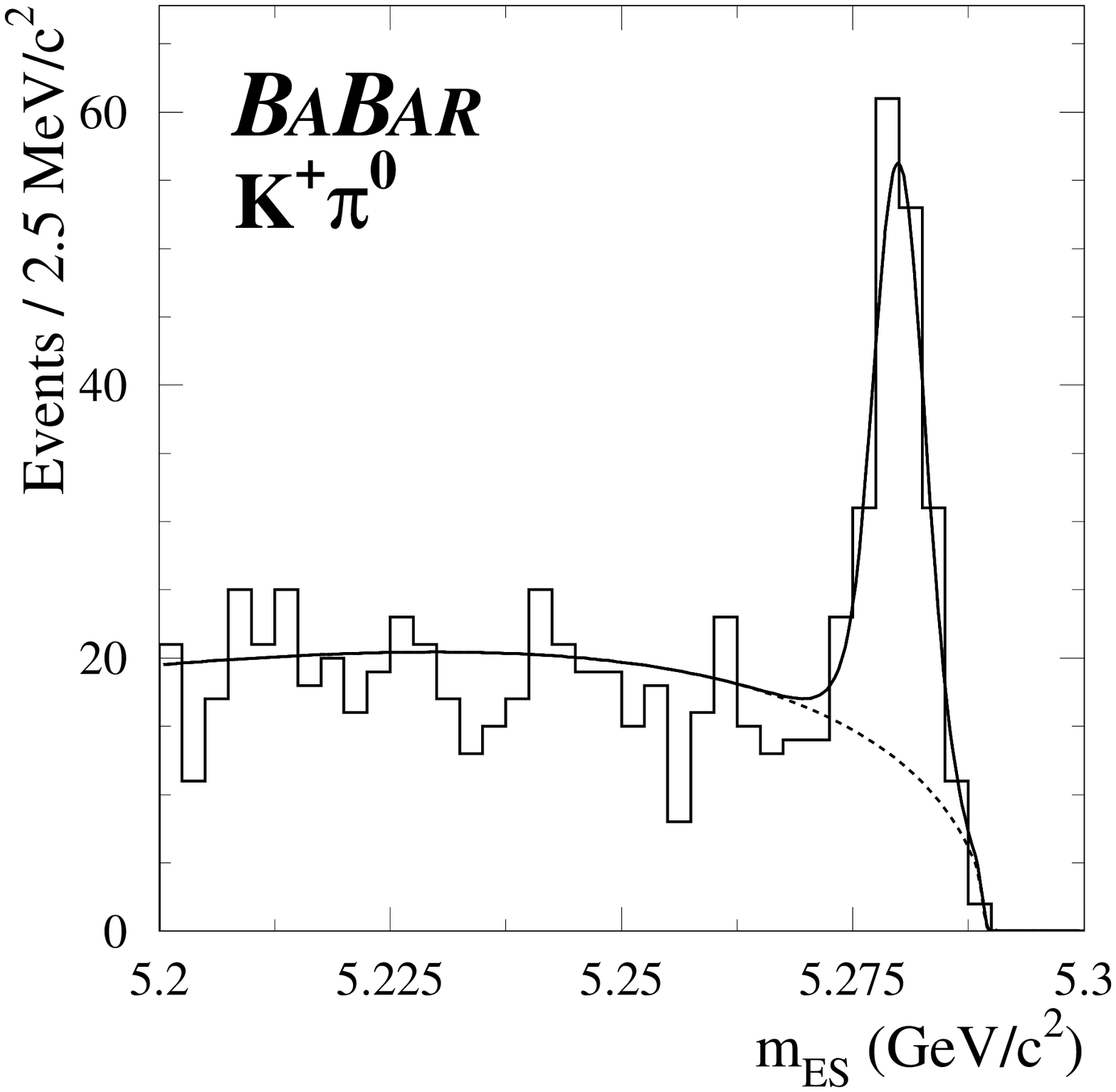}
\end{minipage}
\begin{minipage}[h]{6.0cm}
 \includegraphics[width=6.0cm]{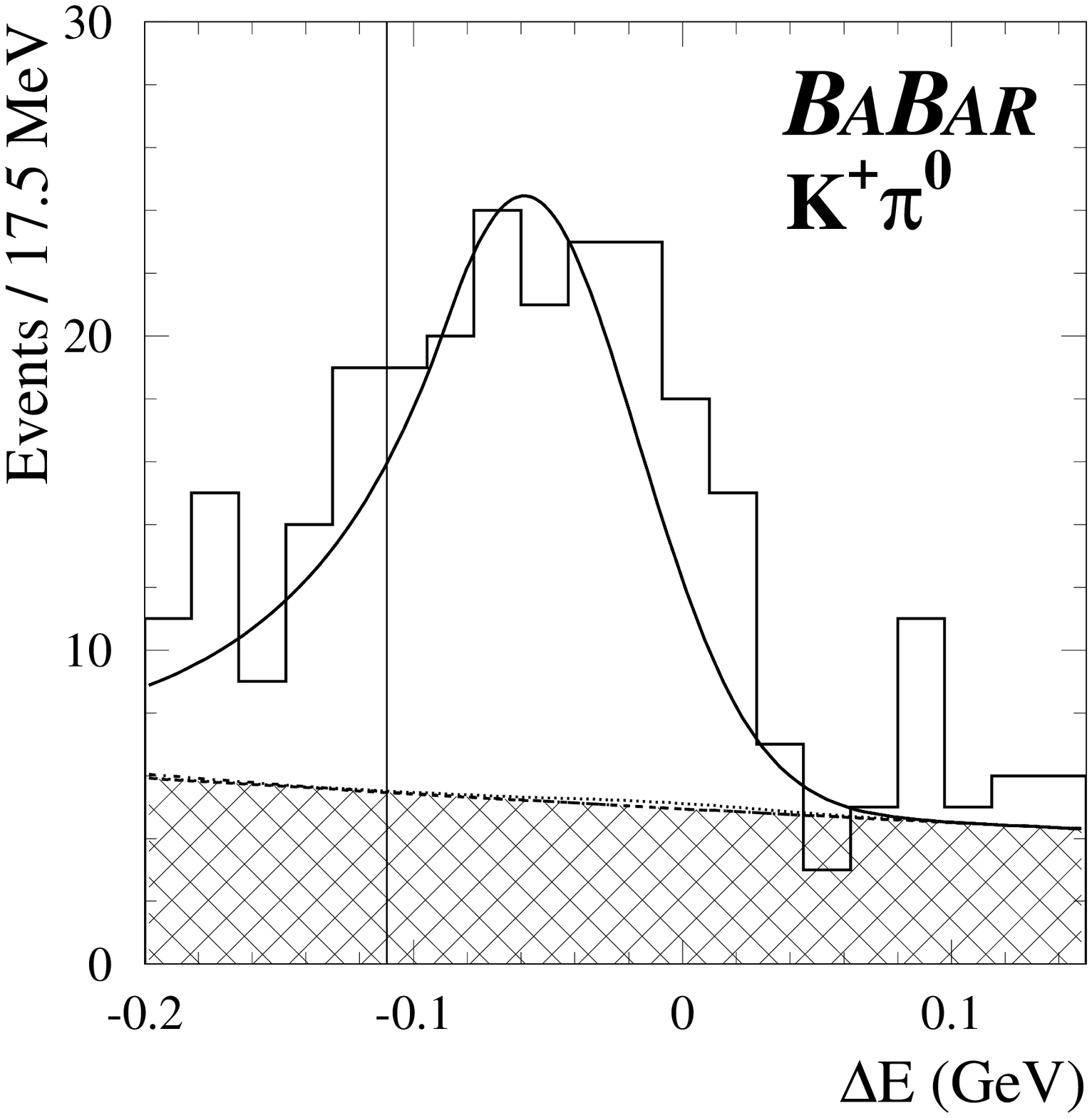}
\end{minipage}
\end{center}
\vspace{-0.9cm}
\begin{center}
\begin{minipage}[h]{6.0cm}
  \includegraphics[width=6.0cm]{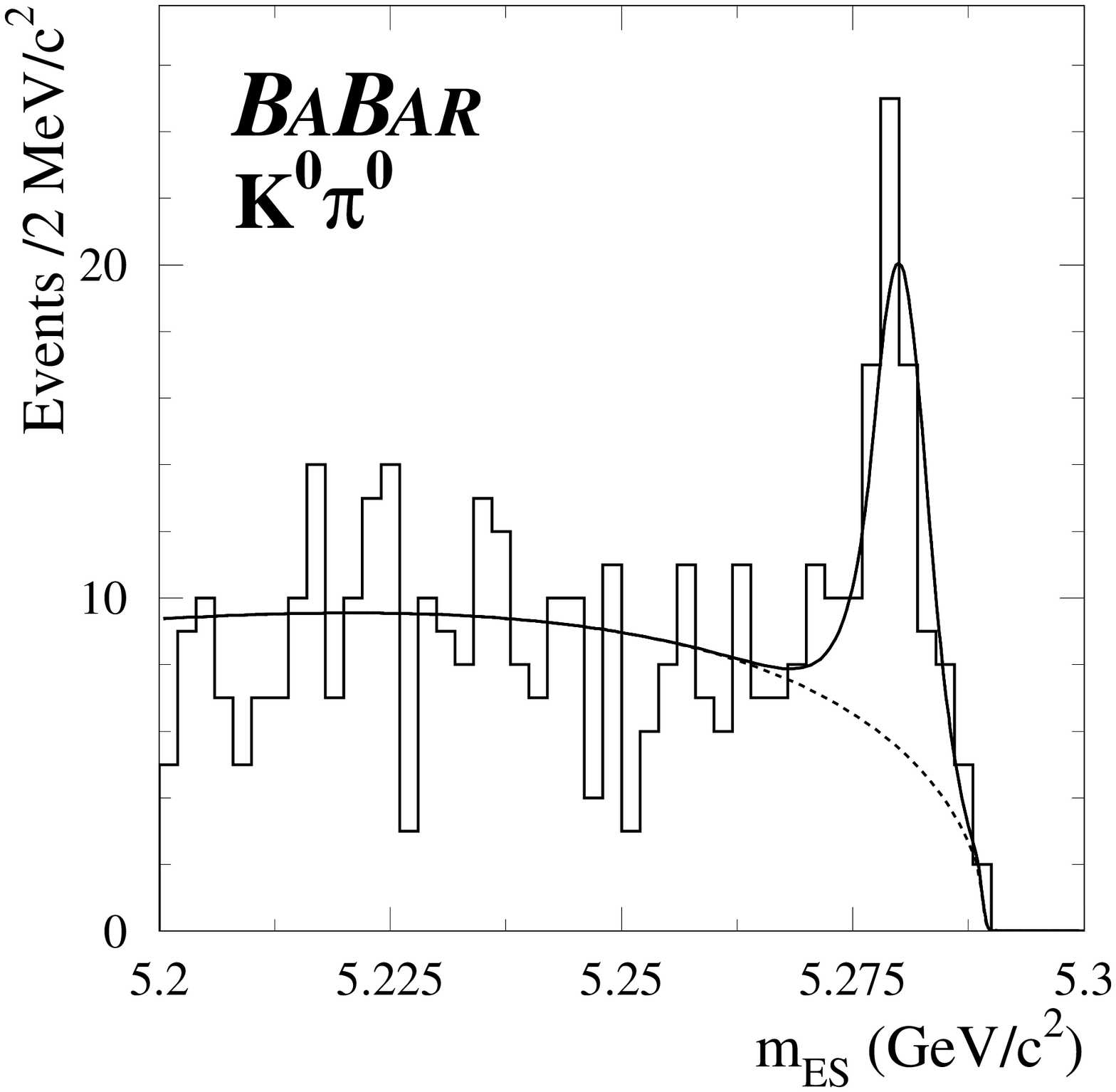}
\end{minipage}
\begin{minipage}[h]{6.0cm}
 \includegraphics[width=6.0cm]{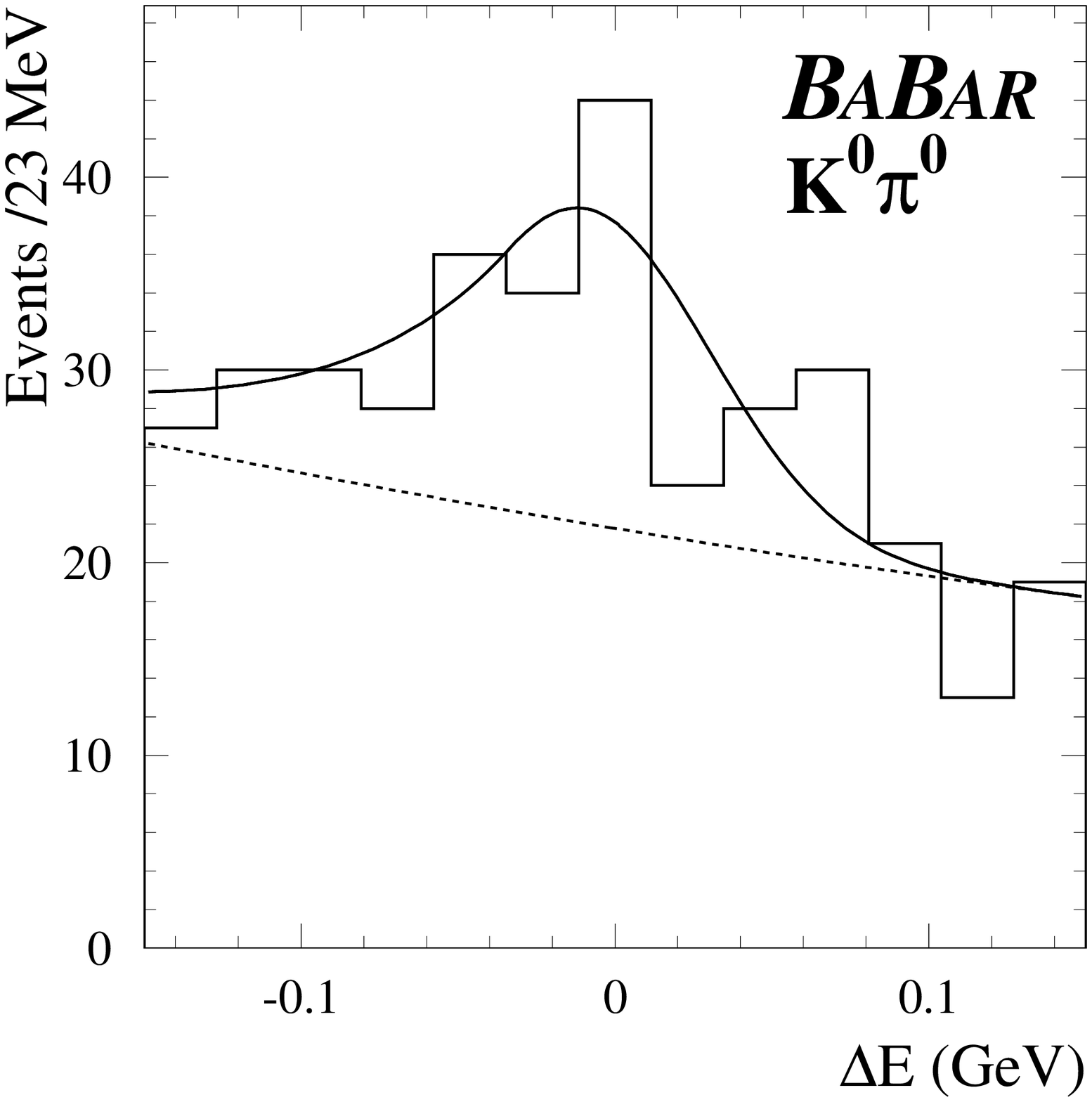}
\end{minipage}
\end{center}
\vspace{-0.9cm}
\caption{Distributions of \mes (left) and $\Delta E$ (right) 
for $\pip \piz$ events (top), $\Kp \piz$ events (center) and
$\Kz \piz$ events (bottom), after additional requirements on
probability ratios, based on all variables except the one being plotted.  
Solid curves represent projections of the complete maximum 
likelihood fit result; dotted curves represent the background 
contribution.
For the \mes\ $\pip \piz$ plot, $\Kp\piz$ cross-feed is also shown
with a dashed curve.
For $\Delta E$ $\pip \piz$ ($\Kp \piz$) plots,
hatched areas represent continuum background only. 
For the $\Delta E$ $\pip \piz$ plot, $\pip\piz$ signal, $\Kp\piz$
cross-feed and $B\Bbar$ background are represented by the dot-dashed,
dotted and dashed curves, respectively.
The $B\Bbar$ background is clearly present at low $\Delta E$ values.
The vertical line represents the $\Delta E$ requirement applied in the
analysis.}
\label{fig:prplots}
\end{figure}

In order to increase the relative fraction of signal events of a given 
type for display purpose we choose events passing requirements on 
probability ratios. These probability ratios are defined as 
${\cal R}_{\rm sig} = \sum_s {\cal P}_s/\sum_i {\cal P}_i$
and ${\cal R}_k = {\cal P}_k/\sum_s {\cal P}_s$,
where $\sum_s$ denotes the sum over the
probabilities for signal hypotheses only, $\sum_i$ denotes the sum 
over all the probabilities (signal and background),
and ${\cal P}_k$ denotes the probability for signal hypothesis $K^+\piz$
(for $h^+ \piz$ only).
These probabilities are constructed from all the PDFs except that 
describing the plotted variable.
Figures \ref{fig:prplots} show the distributions in \mes\  
and $\Delta E $ for events passing all such selection criteria.     
The likelihood fit projections, scaled by the relative efficiencies 
for the probability ratio requirements, are overlaid on each distribution.
Since the sample projections in \mes\ and $\Delta E$ are obtained with 
requirements on different probability ratios, the number of signal events 
appearing in the two projections are not the same.
The efficiencies for the \mes\ plots are: $24\%$, $53\%$ and $65\%$
for $\pip\piz$, $\Kp\piz$ and $\Kz\piz$ signal events, respectively.
For $h^+\piz$ $\Delta E$ projection plots, we show a wider window
($-0.200 < \Delta E < 0.150 \gev$) with respect to the signal region
used in the maximum likelihood fit ($-0.110 < \Delta E < 0.150 \gev$) in
order to show the $B\Bbar$ background present at low $\Delta E$ values.
The efficiencies for the $\Delta E$ plots are: $35\%$, $48\%$ and $98\%$
for $\pip\piz$, $\Kp\piz$ and $\Kz\piz$ signal events, respectively.

\section{Branching Fraction and Direct \boldmath{\CP} Asymmetry Results}
The branching fractions are defined as
\begin{eqnarray}
\label{hpi0br}
{\BR}(h^+\pi^0) &=&  \frac{1}{\BR (\pi^0 \to \gamma \gamma )}
\frac{n_{h^+\pi^0}}{\epsilon_{h^+\pi^0} \cdot N_{\BB}},
\nonumber
\\
\label{kspi0br}
{\BR}(\Kz \piz) &=& \frac{ n_{\KS \piz}}{{\BR} (\pi^0 \to \gamma \gamma )
\cdot {\BR}(K^0 \to \KS) \cdot {\BR}(\KS \to
\pip\pim) \cdot \epsilon_{\KS\piz} \cdot N_{\BB}}\,,
\nonumber
\end{eqnarray}
where $n_{h^+\pi^0}$ ($n_{\KS \piz}$) is the signal yield from the fit
and $\epsilon_{h^+\piz}$ ($\epsilon_{\KS\piz}$) is the reconstruction
efficiency for the mode $h^+\piz$ ($\Kz\piz$) in the detected $\piz$ (\Kz) 
decay chain.
$N_{\BB} = (87.9\pm 1.0)\times 10^6$ is the total number of $\BB$
pairs in our dataset. $\BR(\pi^0 \to \gamma \gamma )$, $\BR(K^0 \to
\KS)$, and $\BR(\KS \to \pip\pim)$ are taken to be equal to
$0.9880$, $0.5$ and $0.686$, respectively~\cite{PDG}. Implicit in
the above equations is the assumption of equal branching fractions for
$\Y4S\to\Bz\Bzb$ and $\Y4S\to\Bu\Bub$.

The \CP\ asymmetry in the $\Kz\piz$ channel is defined as:
\begin{eqnarray}
\label{asym}
\acp_i = \acp_m^i \cdot (1+x_d^2)
\nonumber
\end{eqnarray}
where $x_d = \Delta m_d/\Gamma = 0.755 \pm 0.015$~\cite{PDG}.
Since the flavor of the signal $B$ cannot be determined,
we apply the correction factor $(1+x_d^2)$ to take into
account the $\Bz-\Bzb$ mixing and to translate the measured asymmetry
into the direct $\CP$ asymmetry $\acp_{\CP}$ ignoring \CP-violating
effects in mixing.
In the case of the charged $B$, $\acp_{h^+\piz}=\acp_m^{h^+\piz}$.
Table~\ref{tab:hpi0brresult} summarizes the results on the branching
fractions and the $\CP$ asymmetries: the $90\%$ confidence level (CL)
intervals for the asymmetries are also given.

\begin{table}[!tbp]
\caption{Summary of fitted signal yields, measured branching fraction \BR\,
and $\CP$ asymmetries $\acp_i$. The first error is statistical
and the second is systematic.}
\begin{center}
\smallskip
\begin{tabular}{ccccc} \hline\hline
\smallskip
 Mode & Signal Yield  &
 $\BR\,(10^{-6})$ &  $\acp_i$  & $\acp_i$ ($90\%$ CL) \\ \hline
\smallskip
 $\pip \piz$ & $ 125_{-21}^{+23} \pm 10$ & 
 $5.5 _{-0.9}^{+1.0} \pm 0.6$ & 
 $-0.03 _{-0.17}^{+0.18} \pm 0.02$ & $[-0.32, 0.27]$\\
\smallskip
 $K^+ \piz$ & $239 ^{+21}_{-22}  \pm 6$ 
 & $12.8 _{-1.1}^{+1.2} \pm 1.0$ &
$-0.09 \pm 0.09 \pm 0.01$  & $[-0.24, 0.06]$ \\ 
\smallskip
 $\Kz \piz  $    & $86 \pm 13 \pm 3$     &
$10.4 \pm 1.5 \pm 0.8$   &  
$0.03 \pm 0.36 \pm 0.09$ & $[-0.58, 0.64]$ \\
\hline\hline
\end{tabular}
\end{center}
\label{tab:hpi0brresult}
\end{table}

Systematic uncertainties on the branching fractions arise primarily
from uncertainty on the final selection efficiency and uncertainty
on $n_i$ due to imperfect knowledge of the PDF shapes. The latter is
estimated either by varying the PDF parameters within $1\sigma$ of
their measured uncertainties or by substituting alternative PDFs from
independent control samples. For $\Kz \piz$ analysis, mistag fractions 
and tagging efficiencies are varied by 
$1\sigma$ of their measured uncertainties.

In the $h^+\piz$ analysis the most relevant systematic uncertainties
on the signal yields are due to the Fisher shape for both signal and
background events (about $5\%$ each), 
while for the $\Kz \piz$ analysis the $\Delta E$ background
and Fisher signal shapes contribute with the largest errors. 
We estimate the systematic uncertainty on the signal yields 
due to the residual presence of charmless  backgrounds 
with toy Monte Carlo techniques and we find that it is  
negligible compared with the other effects. 
 
Systematic uncertainties on the $\CP$ asymmetries are evaluated 
from PDF variations. For charged modes, this contribution
is added in quadrature with the limit on intrinsic 
charge bias in the detector (0.01). 
For the $\Kz\piz$ mode, an additional contribution (0.02) coming from
tagging has been evaluated by moving the tagging parameters
(efficiencies and mistag fractions) by $1\sigma$ from
their nominal value.

In conclusion, we have presented preliminary measurements of the branching
fractions of $\Bp \to \pip \piz$, $\Bp \to \Kp \piz$ and $\Bz \to \Kz \piz$.
These measurements supersede our previous results. We do not observe
any evidence of direct \CP\ asymmetry in these channels.

\section{Acknowledgements}
We are grateful for the 
extraordinary contributions of our \pep2\ colleagues in
achieving the excellent luminosity and machine conditions
that have made this work possible.
The success of this project also relies critically on the 
expertise and dedication of the computing organizations that 
support \babar.
The collaborating institutions wish to thank 
SLAC for its support and the kind hospitality extended to them. 
This work is supported by the
US Department of Energy
and National Science Foundation, the
Natural Sciences and Engineering Research Council (Canada),
Institute of High Energy Physics (China), the
Commissariat \`a l'Energie Atomique and
Institut National de Physique Nucl\'eaire et de Physique des Particules
(France), the
Bundesministerium f\"ur Bildung und Forschung and
Deutsche Forschungsgemeinschaft
(Germany), the
Istituto Nazionale di Fisica Nucleare (Italy),
the Research Council of Norway, the
Ministry of Science and Technology of the Russian Federation, and the
Particle Physics and Astronomy Research Council (United Kingdom). 
Individuals have received support from 
the A. P. Sloan Foundation, 
the Research Corporation,
and the Alexander von Humboldt Foundation.

\end{document}